\documentclass[10pt]{article}
\usepackage{graphicx,amsmath,amsthm,amssymb}

\begin{document}
\title{Effects of Noise on Ecological Invasion Processes: Bacteriophage-mediated
Competition in Bacteria
\\
Running head: Effects of Stochastic Noise on Bacteriophage-mediated
Competition in Bacteria}
\author{Jaewook Joo$^{1}$, Eric Harvill$^{2}$, and R\'eka Albert$^{3}$}
\date{\today}
\maketitle
\begin{abstract}
Pathogen-mediated competition, through which an invasive species carrying
and transmitting a pathogen can be a superior competitor to a more vulnerable 
resident species, 
is one of the principle driving forces influencing biodiversity in nature. Using an
experimental system of bacteriophage-mediated competition in bacterial
populations and a deterministic model, we have shown in
Ref.~\cite{joo:2005} that the competitive advantage conferred by the phage
depends only on the relative phage pathology and is independent of the
initial phage concentration and other phage and host parameters such as
the infection-causing contact rate, the spontaneous and infection-induced
lysis rates, and the phage burst size. Here we investigate the effects
of stochastic fluctuations on bacterial invasion facilitated by
bacteriophage, and examine the validity of the deterministic approach. We
use both numerical and analytical methods of stochastic processes to
identify the source of noise and assess its magnitude. We show that the
conclusions obtained from the deterministic model are robust against
stochastic fluctuations, yet deviations become prominently large when the phage are
more pathological to the invading bacterial strain.
\end{abstract}
\footnote[1]{Department of Physics,
Pennsylvania State University, University Park, PA, 16802, USA; 
Tel: (814) 865-4972; Fax: (814) 865-3604; email: jjoo@phys.psu.edu}
\footnote[2]{Department of Veterinary and Biomedical
Science, Pennsylvania State University, University Park, PA, 16802,
USA.}
\footnote[3]{Department of Physics and Huck Institute of the Life Sciences, 
Pennsylvania State University, University Park, PA, 16802, USA.}

\section{INTRODUCTION}

Understanding the ecological implications of infectious disease is one of
a few long-lasting problems that still remains challenging due to their
inherent complexities$^{\cite{anderson:1991,dickmann:2002}}$. 
Pathogen-mediated invasion is one of such ecologically important
processes, where one invasive species carrying and transmitting a pathogen invades into a
more vulnerable species.
Apparent competition$^{\cite{hudson:1998,thomas:2005,bonsall:1997,park:1948}}$, 
the competitive advantage conferred by a pathogen to a less vulnerable species, 
is generally accepted as a major force influencing biodiversity. 
Due to the complexities originating from dynamical interactions among
multiple hosts and multiple pathogens, it
has been difficult to single out and to quantitatively measure the
effect of pathogen-mediated competition in nature. 
For this reason, pathogen-mediated competition and infectious disease dynamics 
in general have been actively studied with theoretical
models$^{\cite{anderson:1991,dickmann:2002,holt:1994,holt:1985,
begon:1992,bowers:1997,greenman:1997}}$.

Theoretical studies of ecological processes generally employ 
deterministic or stochastic modeling approaches. In the former case, the
evolution of a population is described by (partial-) differential or difference
equations$^{\cite{murray:1980}}$. In the latter case, the population is modeled as 
consisting of discrete entities, and its evolution is represented by transition
probabilities. 
The deterministic modeling approach has been favored and widely applied to ecological
processes due to its simplicity and well-established analytic
tools$^{\cite{murray:1980}}$. 
The applicability of the deterministic approach is limited in principle to a system with no
fluctuations and no (spatial) correlations, e.g., a system composed of a large number of
particles under rapid mixing. 
The stochastic modeling approach is more broadly applicable and more comprehensive, 
as the macroscopic equation naturally emerges from a stochastic description of the same
process$^{\cite{vankampen:2001}}$. 
While being a more realistic representation of noisy ecology, the stochastic
approach has a downside: most stochastic models are analytically
intractable and stochastic simulation, a popular alternative, is
demanding in terms of computing time. Nonetheless, the stochastic approach
is indispensable when a more thorough understanding of an ecological
process is pursued.

The role of stochastic fluctuations has been increasingly appreciated 
in various studies of the spatio-temporal patterns of infectious diseases such as
measles$^{\cite{bjornstad:2001}}$, pertussis$^{\cite{rohany:1999}}$ and
syphilis$^{\cite{grassy:2005}}$.
There has been an escalating interest in elucidating the role of stochastic noise not only 
in the studies of infectious disease dynamics but also in other fields 
such as stochastic interacting particle systems as model systems for population
biology$^{\cite{liggett:1999}}$, the stochastic Lotka-Volterra
equation$^{\cite{goel:1971}}$, 
inherent noise-induced cyclic pattern in a predator-prey model$^{\cite{mckane:2005})}$ and 
stochastic gene regulatory
networks$^{\cite{mcadams:1997,hasty:2000,ozbudak:2002,pedraza:2005,
barkai:2001,bialek:2005,vilar:2002}}$.
Here we investigate the effects of noise on pathogen-mediated competition, previously 
only studied by deterministic approaches.

\begin{figure}[hp]
\begin{center}
\includegraphics[width=10cm]{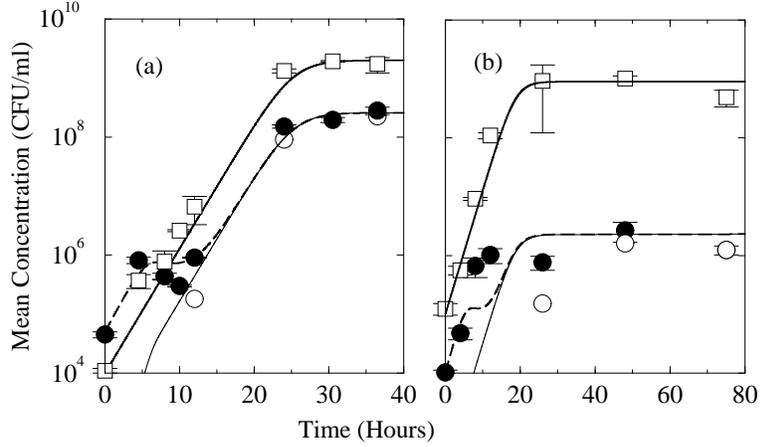}
\caption{\label{fig1} Illustrations of phage-mediated competition 
obtained from {\it in vitro} experiments (symbols) and a deterministic model (lines). 
The phage infection system consists of two genetically identical 
Bordetella bronchiseptica bacteria (Bb) and the bacteriophage BPP-1
($\Phi$)$^{\cite{joo:2005}}$. 
A gentamicin marker (Gm) is used to distinguish the susceptible bacterial strain (BbGm) 
from the phage-carrying bacterial strain (Bb::$\phi$).
As time elapses, a fraction of BbGm become lysogens (BbGm::$\Phi$) due to 
the phage-infection process. Bb::$\Phi$ are represented by open squares and a thick solid line,
BbGm::$\phi$ by open circles and a thin solid line, and the total BbGm 
(BbGm+BbGm::$\Phi$) by filled circles and a long-dashed line, respectively.
(a) Lysogens (Bb::$\Phi$) exogenously and endogenously carrying the prophage invade the BbGm
strain susceptible to phage, and 
(b) lysogens (Bb::$\Phi$) are protected against the invading 
susceptible bacterial strain (BbGm)$^{\cite{joo:2005}}$. 
The differential equations were solved with biologically relevant parameter values. 
(See section 3.1 and Table 1 for a detailed description.)}
\end{center}
\end{figure}

In our previous work$^{\cite{joo:2005}}$ we developed an experimental system and a
theoretical framework for studying bacteriophage-mediated competition in
bacterial populations. 
The experimental system consisted of two genetically identical bacterial
strains; they differed in that one strain was a carrier of the bacteriophage and 
resistant to it while the other strain was susceptible to phage infection. Based on the 
{\it in vitro} experimental set-up, we constructed a differential equation model of
phage-mediated competition between the two bacterial strains. 
Most model parameters were measured experimentally, and a few unknown parameters were
estimated by matching the time-series
data of the two competing populations to the experiments (See
Fig.~\ref{fig1}). The model predicted, and experimental evidence
confirmed, that the
competitive advantage conferred by the phage depends only on the relative phage pathology
and is independent of other phage and host parameters such as the infection-causing contact
rate, the spontaneous and infection-induced lysis rates, and the phage burst size.

Here we examine if intrinsic noise changes the dynamics of the bacterial populations
interacting through phage-mediated competition, and more specifically if it changes the
validity of the conclusions of the deterministic model.
The phage-bacteria infection system is modeled and analyzed with two probabilistic methods:
(i) a linear Fokker-Plank equation obtained by a systematic expansion of a full
probabilistic model (i.e., a master equation), and (ii) stochastic simulations. Both
probabilistic methods are used to identify the source of noise 
and assess its magnitude, through determining the ratio of the standard deviation 
to the average population size of each bacterial strain during the infection process.
Finally stochastic simulations show that the conclusions obtained from 
the deterministic model are robust against stochastic fluctuations, yet 
deviations become large when the phage are more pathological to the invading 
bacterial strain.

\section{A MODEL OF PHAGE-MEDIATED COMPETITION IN BACTERIA}

\begin{figure}[hp]
\begin{center}
\includegraphics[width=10cm]{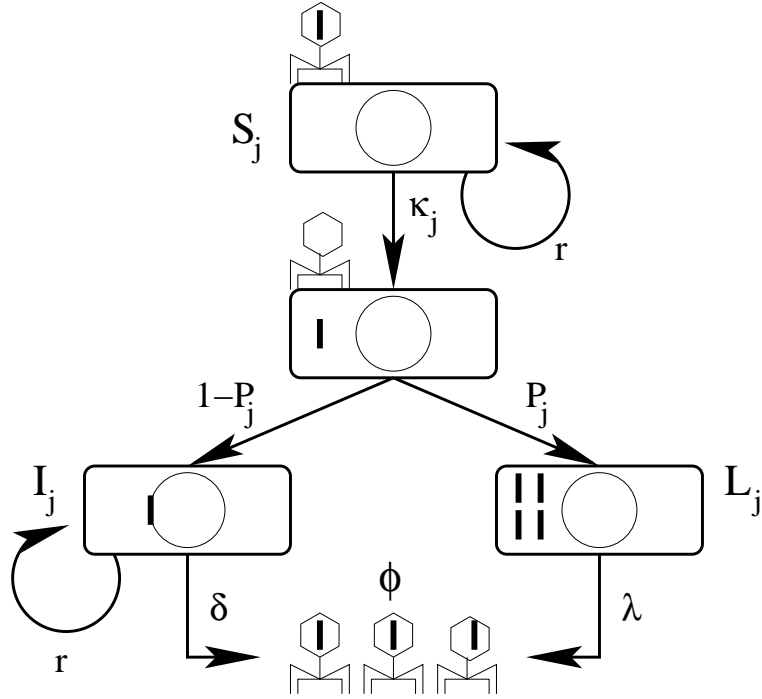}
\caption{\label{fig2} Diagrammatic representation of 
phage-mediated competition between two bacterial strains
with differential susceptibilities $\kappa_j$ and phage pathogenicities 
$P_j$.
The subscript $j \in \{1,2\}$ denotes the type of bacterial strain.
Phage ($\Phi$) are represented by hexagons carrying a thick
segment ($\Phi$ DNA). A susceptible bacterium ($S_j$) is represented
by a rectangle containing an inner circle (bacterial DNA) while a lysogen
($I_j$) is  represented by a rectangle containing $\Phi$ DNA integrated
into its bacterial DNA. All bacterial populations grow with an identical
growth rate $r$ while a latent bacterium ($L_j$) is assumed not to divide.
$\delta$ and $\lambda$ represent spontaneous and infection-induced lysis rates,
respectively.
}
\end{center}
\end{figure}

We consider a generalized phage infection system where two
bacterial strains are susceptible to phage
infection, yet with different degrees of susceptibility 
and vulnerability to phage. The interactions involved in this 
phage-mediated competition between two bacterial strains are 
provided diagrammatically in Fig.~\ref{fig2}. 

We describe this dynamically interacting system with seven homogeneously
mixed subpopulations: Each bacterial strain can be in one of
susceptible ($S_j$), lysogenic ($I_j$), or latent ($L_j$) states,
and they are in direct contact with bacteriophage ($\Phi$). 
All bacteria divide with a constant rate when they are in a log growth phase, 
while their growth is limited when in stationary phase.
Thus we assume that the bacterial population grows with a 
density-dependent rate $r(\Omega)=a(1-\Omega/\Omega_{max})$ 
where $\Omega$ is the total bacterial population and $\Omega_{max}$ is the
maximum number of bacteria supported by the nutrient broth environment. 
Susceptible bacteria ($S_j$) become infected 
through contact with phage at rate $\kappa_j$.
Upon infection the phage can either take a lysogenic pathway or a lytic pathway, 
stochastically determining the fate of the infected bacterium$^{\cite{ptashne:1992}}$.
We assume that a fraction $P_j$ of infected bacteria enter a latent state
($L_j$). Thereafter the phage replicate and then lyse the host bacteria
after an incubation period $1/\lambda$, during which the bacteria
do not divide$^{\cite{ptashne:1992}}$. Alternatively the phage
lysogenize a fraction $1-P_j$ of their hosts, which enter 
a lysogenic state ($I_j$), and incorporate their genome into the DNA of
the host. 
Thus the parameter $P_j$ characterizes the pathogenicity
of the phage, incorporating multiple aspects of phage-host
interactions resulting in damage to host fitness.  
The lysogens ($I_j$) carrying the prophage grow, replicating prophage 
as part of the host chromosome, and are resistant to phage. Even though
these lysogens are very stable$^{\cite{ptashne:1992}}$ without
external perturbations, spontaneous induction can occur at a low
rate $\delta$, consequently replicating the phage and lysing the
host bacteria. In general, both the number of phage produced
(the phage burst size $\chi$ ) and the phage pathology $P_j$ depend on the
culture conditions$^{\cite{ptashne:1992}}$. The two bacterial strains
differ in susceptibility ($\kappa_j$) and vulnerability ($P_j$) 
to phage infection.

When the initial population size of the invading bacterial strain is small, 
the stochastic fluctuations of the bacterial population size are expected to be large
and likely to affect the outcome of the invasion process.
A probabilistic model of the phage infection system is able to capture the 
effects of intrinsic noise on the population dynamics of bacteria. 
Let us define the joint probability density
$P_t(\underline{\eta})$ denoting the probability of the system to
be in a state $\underline{\eta}(t)=(S_1,I_1,L_1,S_2,I_2,L_2,\Phi)$
at time $t$ where $S_j$, $I_j$ and $L_j$ denote the number of bacteria
in susceptible, latent or infected states, respectively. 
The time evolution of the joint probability is
determined by the transition probability per unit time 
$T(\underline{\eta'}|\underline{\eta};t)$ of going from a state
$\underline{\eta}$ to a state $\underline{\eta'}$.
We assume that the transition probabilities do not depend on the history of the previous
states of the system but only on the immediately past state. 
There are only a few transitions that are allowed to take place. 
For instance, the number of susceptible bacteria increases
from $S_1$ to $S_1+1$ through the division of a single susceptible bacterium 
and this process takes place with the transition rate
$T(S_1+1,I_1,L_1,S_2,I_2,L_2,\Phi|S_1,I_1,L_1,S_2,I_2,L_2,\Phi)$=$r(\Omega) S_1$.
The allowed transition rates are
\begin{eqnarray}
\label{transition}
&&T(S_j+1,...|S_j,...;t)= r(\Omega) S_j,
\\ \nonumber
&&T(...,I_j+1,...|...,I_j,...;t)= r(\Omega) I_j,
\\ \nonumber
&&T(S_j-1,I_j+1,...,\Phi-1|S_j,I_j,...,\Phi;t) =\kappa_j (1-P_j) \Phi S_j
\\ \nonumber
&&T(S_j-1,...,L_j+1,\Phi-1|S_j,...,L_j,\Phi;t) = \kappa_j P_j \Phi S_j
\\ \nonumber
&&T(...,I_j-1,...,\Phi+\chi|...,I_j,...,\Phi;t)= \delta I_j
\\ \nonumber
&&T(...,L_j-1,...,\Phi+\chi|...,L_j,...,\Phi;t)= \lambda L_j
\end{eqnarray}
where $\Omega(t)=\sum_j (S_j(t)+I_j(t)+L_j(t))$.
The second line represents the division of a lysogen;
the 3rd line describes an infection process by phage taking a lysogenic pathway 
while the fourth line denotes an infection process by phage taking a lytic pathway. 
The last two transitions are spontaneously-induced and infection-induced lysis
processes, respectively.
Bacterial subpopulations that are unchanged during a particular transition are denoted by
``$...$''. 
The parameters $a$, $k_j$, $\delta$ and $\lambda$ in the transition rates of 
Eq.~(\ref{transition}) represent the inverse of the expected waiting
time between events in an exponential event distribution and they are equivalent to
the reaction rates given in Fig.~\ref{fig2}.

The stochastic process specified by the
transition rates in Eq.~(\ref{transition}) is Markovian, thus we can immediately
write down a master equation governing the time evolution of the 
joint probability $P(\underline{\eta})$. 
The rate of change of the joint probability $P_t(\underline{\eta})$ is the
sum of transition rates from all other states $\underline{\eta'}$ to
the state $\underline{\eta}$, minus the sum of transition rates from
the state $\underline{\eta}$ to all other states
$\underline{\eta'}$:
\begin{eqnarray}
\label{master}
\frac{d P_t(\underline{\eta})}{d t} &=& \sum_j \Bigl \{
(E^{-1}_{S_j}-1)[T(S_j+1,...|S_j,...;t)P_t(\underline{\eta})
\\ \nonumber
&+&
(E^{-1}_{I_j}-1)[T(...,I_j+1,...|...,I_j,...;t)P_t(\underline{\eta})]
\\ \nonumber
&+&
(E^{+1}_{\Phi}E^{+1}_{S_j}E^{-1}_{I_j}-1)[T(S_j-1,I_j+1,...,\Phi-1|S_j,I_j,...,\Phi;t)
P_t(\underline{\eta})]
\\ \nonumber
&+&(E^{+1}_{\Phi}E^{+1}_{S_j}E^{-1}_{L_j}-1)[T(S_j-1,...,L_j+1,\Phi-1|S_j,...,L_j,\Phi;t)
P_t(\underline{\eta})]
\\ \nonumber
&+& \delta(E^{+1}_{I_j}E^{-\chi}_{\Phi}-1)[T(...,I_j-1,...,\Phi+\chi|...,I_j,...,\Phi;t)
P_t(\underline{\eta})]
\\ \nonumber
&+& \lambda (E^{+1}_{L_j}E^{-\chi}_{\Phi}-1)T(...,L_j-1,...,\Phi+\chi|...,L_j,...,\Phi;t)
P_t(\underline{\eta})]
\Bigr \}
\nonumber
\end{eqnarray}
where $E^{\pm1}_{\alpha}$ is a step operator which acts on
any function of $\alpha$ according to
$E^{\pm1}_{\alpha}f(\alpha,...)=f(\alpha \pm 1,...)$.

The master equation in Eq.~(\ref{master}) is nonlinear and
analytically intractable. There are two alternative ways to seek a
partial understanding of this stochastic system: a stochastic simulation 
and a linear Fokker-Plank equation obtained from a systematic 
approximation of the master equation.  
A stochastic simulation is one of the most accurate/exact methods to study the
corresponding stochastic system. However, stochastic simulations of 
an infection process in a large system are very demanding
in terms of computing time, even today. 
Moreover, simulation studies can explore only a relatively small fraction of 
a multi-dimensional parameter space, thus provide neither 
a complete picture nor intuitive insight to the current infection 
process. 
The linear Fokker-Plank equation is only an approximation of the full
stochastic process; it describes the time-evolution of the probability density, 
whose peak is moving according to macroscopic equations. In cases
where the macroscopic equations are nonlinear, one needs to go
beyond a Gaussian approximation of fluctuations, i.e., the higher moments 
of the fluctuations should be considered. In cases when an analytic solution 
is possible, the linear Fokker-Plank equation method can overcome most
disadvantages of the stochastic simulations. Unfortunately such
an analytic solution could not be obtained for the master equation in Eq.~(\ref{master}).

In the following sections we present a systematic expansion method of 
the master equation to obtain both the macroscopic equations and the linear Fokker-Plank
equation, then an algorithm of stochastic simulations.

\section{SYSTEMATIC EXPANSION OF THE MASTER EQUATION}

In this section we will apply van Kampen's elegant method$^{\cite{vankampen:2001}}$ 
to a nonlinear stochastic process, in a system whose size increases exponentially
in time. 
This method not only allows us to obtain a deterministic version of the stochastic 
model in Eq.~(\ref{master}) but also gives a method of finding stochastic corrections to the
deterministic result. We choose an initial system size 
$\Omega_o=\sum_j (S_j(0)+I_j(0)+L_j(0))+\Phi(0)$ 
and expand the master equation in order of $\Omega^{-1/2}_o$.
We do not attempt to prove the validity of our application
of van Kampen's $\Omega_o$-expansion method to this nonlinear stochastic system; a
required condition for valid use of $\Omega_o$-expansion scheme, namely the 
stability of fixed points, is not satisfied because the system size increases indefinitely
and there is no stationary point.
However, as shown in later sections, the linear Fokker-Plank equation 
obtained from this $\Omega_o$-expansion 
method does provide very reliable results, comparable to the results of 
stochastic simulations.

In the limit of infinitely large $\Omega_o$, the variables ($S_j$,
$I_j$, $L_j$, $\Phi$) become deterministic and equal to ($\Omega_o s_j,
\Omega_o i_j, \Omega_o l_j, \Omega_o \phi$), where ($s_j,i_j,l_j,\phi$)
are normalized quantities, e.g., $s_j=S_j/\Omega_o$. In this infinitely large size limit
the joint probability $P_t(\underline{\eta})$ will be a delta function
with a peak at ($\Omega_o s_j, \Omega_o i_j, \Omega_o l_j, \Omega_o \phi$).
For large but finite $\Omega_o$, we would expect $P(\underline{\eta})$ to have a
finite width of order $\Omega^{1/2}_o$. 
The variables ($S_j$,
$I_j$, $L_j$, $\Phi$) are once again stochastic and we
introduce new stochastic variables
($\xi_{S_j},\xi_{I_j}, \xi_{L_j},\xi_{\Phi}$):
$S_j=\Omega_o s_j+\Omega^{\frac{1}{2}}_o \xi_{S_j}$,
$I_j=\Omega_o i_j+\Omega^{\frac{1}{2}}_o \xi_{I_j}$,
$L_j=\Omega_o l_j+\Omega^{\frac{1}{2}}_o \xi_{L_j}$,
$\Phi_j=\Omega_o \phi_j+\Omega^{\frac{1}{2}}_o \xi_{\Phi_j}$.
These new stochastic variables represent inherent noise and contribute to
deviation of the system from the macroscopic dynamical behavior.

The new joint probability density function $\Pi_t$ is defined by
$P_t(\underline{\eta})=\Pi_t(\underline{\xi})$ where
$\underline{\xi}=(\xi_{S_1},\xi_{I_1},\xi_{L_1},\xi_{S_2},\xi_{I_2},\xi_{L_2},\xi_{\Phi})$.
Let us define the step operators $E^{\pm}_{\alpha}$, which change $\alpha$ into $\alpha \pm 1$
and therefore $\xi_\alpha$ into $\xi_\alpha + \Omega^{-1/2}_o$, so that in new variables
\begin{equation}
E^{\pm 1}_{\alpha} =1\pm \Omega^{-\frac{1}{2}}_o\frac{\partial}{\partial \xi_{\alpha}}
+\frac{\Omega^{-1}_o}{2}\frac{\partial^{2}}{\partial \xi_{\alpha}^{2}} \pm ...
\end{equation}
The time derivative of the joint probability $P_t(\underline{\eta})$ 
in Eq.~(\ref{master}) is taken at a fixed state
$\underline{\eta}=(S_1,I_1,L_1,S_2,I_2,L_2,\Phi)$, 
which implies that the time-derivative taken on both sides of
$\alpha=\Omega_o \alpha + \Omega^{1/2}_o \xi_{\alpha}$ should lead to
$d \xi_{\alpha}/dt=-\Omega^{1/2}_o d \alpha/dt$ where $\alpha$ can be 
either $S_1$, $I_1$, $L_1$, $S_2$, $I_2$, $L_2$, or $\Phi$. Hence,
\begin{equation}
\frac{d P(\underline{\eta};t)}{dt}= \frac{\partial
\Pi(\underline{\xi})}{\partial t}
-\sum_{\alpha=S_1,S_2,I_1,I_2,L_1,L_2,\Phi} \Bigl \{
\Omega^{\frac{1}{2}}_o \frac{\partial \alpha}{dt} \frac{\partial
\Pi(\underline{\xi};t)}{\partial \xi_{\alpha}} \Bigr \}.
\end{equation}
We shall assume that the joint probability density is a delta function at
the initial condition $\underline{\eta_o}$, i.e.,
$P_0(\underline{\eta})=\delta_{\underline{\eta},\underline{\eta_o}}$.

The full expression of the master equation in the new variables is 
shown in appendix A. Here we collect several
powers of $\Omega_o$. In section 3.1 we show that macroscopic
equations emerge from the terms of order $\Omega^{1/2}_o$ and that 
a so-called invasion criterion, defined as the condition 
for which one bacterial population outcompetes the other, can
be obtained from these macroscopic equations.
In section 3.2 we show that the terms of order $\Omega^{0}_o$
give a linear Fokker-Plank equation whose time-dependent coefficients 
are determined by the macroscopic equations.

\subsection{Emergence of the Macroscopic Equations}

There are a few terms of order $\Omega^{1/2}_o$ in the master
equation in the new variables as shown in appendix A, 
which appear to make a large $\Omega_o$-expansion of the master 
equation improper.  
However those terms in order of $\Omega^{1/2}_o$ cancel 
if the following equations are satisfied
\begin{eqnarray}
\label{macroscopic} \frac{d s_j}{dt}&=& r(\Omega) s_{j}-\kappa_j
\Omega_o \phi s_j
\\ \nonumber
\frac{d i_j}{dt}&=& (1-P_j)\kappa_j \Omega_o \phi s_j +
(r(\Omega)-\delta) i_j
\\  \nonumber
\frac{d l_j}{dt}&=& P_j \kappa_j \Omega_o \phi s_j -\lambda l_j
\\  \nonumber
\frac{d \phi}{dt}&=& \chi \sum_j (\delta \phi_j + \lambda
l_j)-\sum_j \kappa_j \Omega_o \phi s_j \nonumber
\end{eqnarray}
Eq.~(\ref{macroscopic}) are identical to the deterministic equations 
of the corresponding stochastic model in
the limit of infinitely large $\Omega_o$, i.e., in the limit of
negligible fluctuations.

These equations allow for the derivation of the invasion criterion, 
defined as the choice of the system parameters in Table 1 that makes one invading bacterial 
strain dominant in number over the other strain. 
Suppose that an initial condition of Eq.~(\ref{macroscopic})
is $s_1(0)>0$, $s_2(0)>0$, $i_1(0)>0$,
$\phi(0) \ge 0$, and $i_2(0)=l_1(0)=l_2(0)=0$.
(a) In the case of $\phi(0)=\delta=0$, there is no phage-mediated
interaction between bacteria and the ratio of
$s_1(t):s_2(t):i_1(t)$ remains unchanged for $t \ge 0$.
(b) However when either ($\phi(0)=0$ and $\delta>0$) or $\phi(0)>0$,
the above ratio changes in time due to phage-mediated interactions.
Even though in principle these nonlinear coupled equations are unsolvable,
we managed to obtain an analytic solution of the macroscopic 
Eq.~(\ref{macroscopic}) in the limit of a fast infection process, 
i.e., ($\kappa_j \Omega_o s_2(0)/a >>1$ and $\lambda/a>>1$),
by means of choosing appropriate time-scales and using
a regular perturbation theory$^{\cite{murray:1980}}$.
(See Ref.~\cite{joo:2005} for a detailed description in a simpler system.)
We found a simple relationship between the ratios of 
the two total bacterial populations:
\begin{equation}
\label{invasion}
r_{12}(t)=r_{12
}(0)(1-P_1)/(1-P_2)
\end{equation}
where $r_{12}(t) \equiv \frac{s_1(t)+i_1(t)+l_1(t)}{s_2(t)+i_2(t)+l_2(t)}$ 
for a sufficiently long time $t$.
Thus the final ratio $r_{12}(t)$ is determined solely by three quantities, 
the initial ratio, $r_{12}(0)$, and the two phage pathologies, and
is independent of other kinetic parameters such as 
the infection-causing contact rate, the 
spontaneous and infection-induced lysis rates, and the phage burst size. 
The invasion criterion, the condition for which bacterial strain 1 outnumbers 
bacterial strain 2, is simply $r_{12}(t)>1$.

\begin{figure}
\begin{center}
\includegraphics[width=10cm]{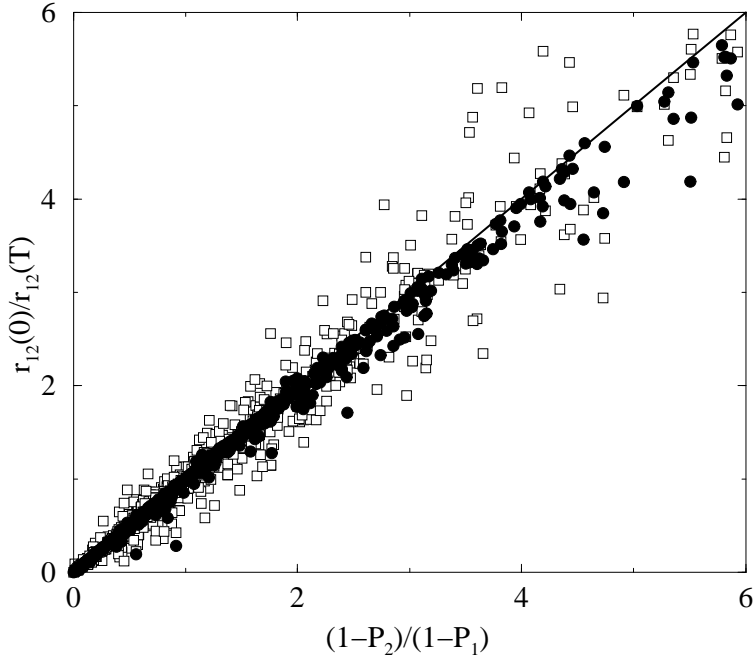}
\caption{\label{fig3} Numerical verification of the invasion criterion of
Eq.~(\ref{invasion}) for a generalized deterministic infection system 
where both bacterial strains are susceptible to phage infection.
The ratio $r_{12}(0)/r_{12}(T)$ was numerically evaluated by
solving Eq.~(\ref{macroscopic}) with 2000 sets of parameters chosen 
uniformly in the intervals $0<P_1,P_2<1$ for phage pathologies, 
$1/min\{P_1,P_2\}<\chi<100$ for the phage burst size,
$0<\lambda/a<0.5$ for the spontaneous induction rate,
$10^{-1}s_2(0)<s_1(0)<10 s_2(0)$ and $0<i_1(0),\phi(0)<10^{-2}
s_2(0)$ for the initial concentrations of bacterial strains and phage. 
The time $T$ is chosen to be a sufficiently long time.
Filled circles represent the data from 1000
sets of parameters with relatively large $\Omega_o \kappa_j s_j/a$
and $\lambda/a$ (e.g., $0.1<\Omega_o \kappa_j s_j/a, \lambda/a <10$).
Open circles are from another 1000 sets of parameters with small
$\Omega_o \kappa_j s_j/a$ and $\lambda/a$ (e.g., $0<\Omega_o \kappa_j
s_j/a ,\lambda/a<0.1$). }
\end{center}
\end{figure}

To validate the invasion criterion of Eq.~(\ref{invasion})
in the range of relatively small values of $\kappa_{j} \Omega_o s_2(0)/a$ and $\lambda/a$,
we solved Eq.~(\ref{macroscopic}) numerically with 2000 sets of parameters selected randomly
from the biologically relevant intervals. 
Fig.~\ref{fig3} shows that the simple relationship in Eq.~(\ref{invasion}) 
between $r_{12}(0)/r_{12}(t)$ and $(1-P_2)/(1-P_1)$ is robust against parameter 
variations. The results deviate from the linear relationship with increasing phage
pathology on the invading bacterial strain 1 compared with that on bacterial strain 2, 
i.e., $(1-P_2)/(1-P_1)>>1$, or $P_1>>P_2$.

\subsection{Linear Noise Approximation: a Linear Fokker Plank Equation}

For simplicity we will assume hereafter that all bacteria grow 
with a growth rate $r=a$ in a log phase, i.e., there is no resource competition. 
Identifying terms of $\Omega^{0}_o$ in the power expansion 
of the master equation (see appendix A) 
we obtain a linear Fokker-Plank equation (see appendix B). 
This approximation is called as linear noise approxiamtion~\cite{vankampen:2001}
and the solution of the linear Fokker-Plank equation in appendix A is a
Gaussian~\cite{vankampen:2001}, which
means that the probability distribution
$\Pi_t(\underline{\xi})$ is completely specified by the first two
moments, $\langle \xi_{\alpha}(t) \rangle$ and $\langle
\xi^{2}_{\alpha}(t) \rangle$, where $\alpha=S_j, I_j, L_j, \Phi$.

Multiplying the Fokker-Plank equation by $\xi_{\alpha}$ and
$\xi_{\alpha}\xi_{\alpha'}$ and integrating over all
$\underline{\xi}$ we find the time-evolution of the first and
the second moments of noise, $\langle \xi_\alpha \rangle$ and 
$\langle \xi_{\alpha} \xi_{\alpha'} \rangle$ (see appendix C). 
The solutions of all first moments
are simple: $\langle \xi_{\alpha}(t) \rangle=0$ for
all $t$, provided that the initial condition is chosen such that
initial fluctuations vanish, i.e., $\langle \xi_{\alpha}(0)
\rangle=0$. 
The differential equations governing the time evolution of 
the second moments are coupled, and their solutions can only be 
attained by means of numerical integrations. 
We use the time evolution of the second moments of noise to study  
the role of stochastic fluctuations on phage-mediated competition, 
and especially to investigate the effects of noise on the invasion criterion.  
Let $\delta N_j$ be the deviation of the total population size $N_j$ of the $j$th 
bacterial strain from its average value,
i.e., $\delta N_j=N_j-\langle N_j \rangle=\Omega^{1/2}_o(\xi_{S_j}+\xi_{I_j}+\xi_{L_j})$ 
where $N_j=S_j+I_j+L_j$ and 
$\langle N_j \rangle$=$\langle S_j \rangle+\langle I_j \rangle$+$\langle L_j \rangle$.
Let us define the normalized variance of the total population size 
of the $j$th bacterial strain 
\begin{equation}
Var(N_j) \equiv \frac{\langle \delta N_j^{2} \rangle }{\langle  N_j \rangle^{2} } 
=\frac{\Omega_o}{\langle N_j \rangle^{2}} \Bigl \{ \langle
\xi_{S_j}^{2} \rangle + \langle \xi_{I_j}^{2} \rangle + \langle
\xi_{L_j}^{2} \rangle +2( \langle \xi_{S_j} \xi_{I_j} \rangle
+\langle \xi_{S_j} \xi_{L_j} \rangle +\langle \xi_{I_j} \xi_{L_j}
\rangle ) \Bigr \}
\end{equation}
where $\langle . \rangle$ is a statistical ensemble average. 
The square root of the normalized variance, 
$\sqrt{\langle \delta N_j^{2}(t) \rangle}/\langle N_j(t) \rangle$, is the 
magnitude of noise of the $j$th bacterial strain at a given time t. 
Another useful quantity is the normalized co-variance between 
the $i$th bacterial strain in a state $\alpha$ and the $j$th bacterial strain 
in a state $\beta$: 
\begin{equation}
Cov(\alpha_i,\beta_j) \equiv \frac{\langle \delta \alpha_i \delta \beta_j \rangle}
{\langle N_i \rangle \langle N_j \rangle}
=\frac{\Omega_o \langle \xi_{\alpha_i} \xi_{\beta_j}\rangle}
{\langle N_i \rangle \langle N_j \rangle} 
\end{equation} 
We will present the results for these variances and co-variances in section 5.

\section{THE GILLESPIE ALGORITHM FOR STOCHASTIC SIMULATIONS}

In this section we briefly describe our application of the Gillespie 
algorithm$^{\cite{gillespie:1977}}$ for simulation of 
the stochastic process captured in the master equation of Eq.~(\ref{master}), 
where in total 12 biochemical reactions take place stochastically.
The Gillespie algorithm consists of the iteration of the following steps:
(i) selection of a waiting time $\tau$ during which no reaction occurs,
\begin{equation}
\tau=-\frac{1}{\sum_{j} a_j} ln \theta
\end{equation}
where $\theta$ is a random variable uniformly chosen from an
interval $(0,1)$ and $a_j$ is the reaction rate for the $j$th
biochemical reaction. 
(ii) After such a waiting time, which
biochemical reaction will take place is determined by the following algorithm.
The occurrence of each event has a weight $a_j/\sum_j a_j$. 
Thus the $i$th biochemical reaction
is chosen if $\sum_{j=1}^{i} a_j< \theta' \sum^{N}_{j=1}
a_j<\sum_{j=1}^{i+1} a_j$ where $\theta'$ is another random number
selected from the interval $(0,1)$ and $N$ is the total number of biochemical reactions. 
(iii) After execution of the $j$th reaction, 
all reaction rates that are affected by the $j$th reaction are updated.

We measure the averages, the normalized variances and co-variances of bacterial populations 
at various time points, 
by taking an average over $10^{4}$ realizations of the infection process, starting 
with the same initial condition.
Because a normalized variance or covariance is a measure of deviations of a stochastic variable 
from a macroscopic value (which is regarded as a true value), 
it is not divided by the sampling size.

The computing time of the Gillespie algorithm-based simulations increases exponentially
with the system size.
In the absence of resource competition, the total bacterial population increases
exponentially in time. 
Because we need to know the stationary ratio of the two bacterial populations,
the computing time should be long enough compared to typical time scales of 
the infection process. 
This condition imposes a limit on the range of parameters that we can explore
to investigate the validity of the invasion criterion. We choose the values of parameters from 
the biologically relevant range given in Table 1 and we, furthermore, set lower bounds on the rates 
of infection causing contact $\kappa_j$ and infection-induced lysis $\lambda$, 
namely $\kappa_j>\kappa_o$ and $\lambda>\lambda_o$.

\section{RESULTS}

While the methodologies described in section 3 and 4 apply to the 
general case of two susceptible bacterial strains, in this section 
we limit our investigations to a particular infection
system, called a ``complete infection system'' hereafter, 
in which bacterial strain 1 is completely lysogenic and only bacterial strain 2  
is susceptible to phage infection. 
There are two advantages to studying the complete infection system: 
1) this is equivalent to the infection system which we studied
experimentally$^{\cite{joo:2005}}$ and thus the results are immediately 
applicable to at least one real biological system.
2) the probabilistic description of bacterial strain 1 (lysogens) 
is analytically solvable as it corresponds to a stochastic birth-death 
process$^{\cite{vankampen:2001}}$. 
In section 5.1, studying a system consisting of only lysogens, 
we elucidate the different dynamic patterns of the normalized variance
when the system size remains constant or when it increases. 
This finding provides us with the asymptotic behavior of the 
normalized variances of both bacterial strains 
because both strains become lysogens eventually after all susceptible bacteria are depleted from the system.
In section 5.2, we investigate the role of stochastic noise on phage-mediated competition by identifying 
the source of noise and assessing its magnitude in the complete infection system.
Finally in section 5.3, we investigate the effect of noise on the invasion criterion by means of stochastic simulations.

\subsection{Stochastic Birth-death Process: Growth and Spontaneous Lysis of Bacterial
Strain 1}
The dynamics of lysogens of bacterial strain 1 is completely decoupled
from that of the rest of the complete infection system and can be studied independently. 
They grow at a rate $r$ and are lysed at a rate $\delta$.
There exists an exact solution for the master equation of this stochastic 
birth-death process. 
Thus we can gauge the accuracy of an approximate method for the corresponding 
stochastic process by comparing it with the exact solution. 
(See appendix C for description of the birth-death process 
and its exact solution.) 
The master equation of the birth-death process is
\begin{equation}
\label{master-BD} \frac{dP_t(I_1)}{dt}=(E^{-1}_{I_1}-1)r I_1
P_t(I_1)+(E^{+1}_{I_1}-1)\delta I_1 P_t(I_1)
\end{equation}
where $I_1(t)$ represents the number of lysogens at time $t$.
$I_1$ is transformed into a new variable $\xi_{I_1}$ as discussed 
in section 3, which results in $I_1=\Omega_o i_1 +\Omega^{1/2} \xi_{I_1}$, $P_t(I_1)=\Pi_t(\xi_{I_1})$, 
and $E^{\pm 1}_{I_1}=1 \pm \Omega^{-1/2}_o \frac{\partial}{\partial 
\xi_{I_1}}+\frac{\Omega^{-1}_{o}}{2}\frac{\partial^{2}}{\partial \xi^{2}_{I_1}}$. 
Then keeping terms of order $\Omega^{0}_{o}$ from $\Omega_o$-expansion of Eq.~(\ref{master-BD}), 
we obtain the linear Fokker-Plank
equation,
\begin{equation}
\label{BD-FP}
\frac{\partial \Pi_t(\xi_{I_1})}{\partial t}=(r+\delta)\frac{i_{1}}{2}
\frac{\partial^{2} \Pi_t(\xi_{I_1})}{\partial \xi^{2}_{I_1}}
+(\delta-r)\frac{\partial \xi_{I_1} \Pi_t(\xi_{I_1})}{\partial \xi_{I_1}}
\end{equation}
where $i_1(t)=I_1(t)/\Omega_o$ is a normalized quantity that evolves according to
$\frac{d i_1(t)}{dt}=(r-\delta)i_1(t)$ and $\Omega_o=I_1(0)$.
Multiplying by $\xi_{I_1}$ and $\xi^{2}_{I_1}$ both sides of Eq.~(\ref{BD-FP}) and
integrating over $\xi_{I_1}$, we obtain the equations for the
first and the second moments of noise $\xi_{I_1}$:
\begin{eqnarray}
\label{BD-moments}
\frac{d \langle
\xi_{I_1}\rangle}{dt}&=&(r-\delta) \langle \xi_{I_1} \rangle
\\ \nonumber
\frac{d \langle \xi^{2}_{I_1}\rangle}{dt}&=&(r+\delta) i_1
+2(r-\delta)\langle \xi^{2}_{I_1}\rangle
\end{eqnarray}

\begin{figure}
\begin{center}
\includegraphics[width=10cm]{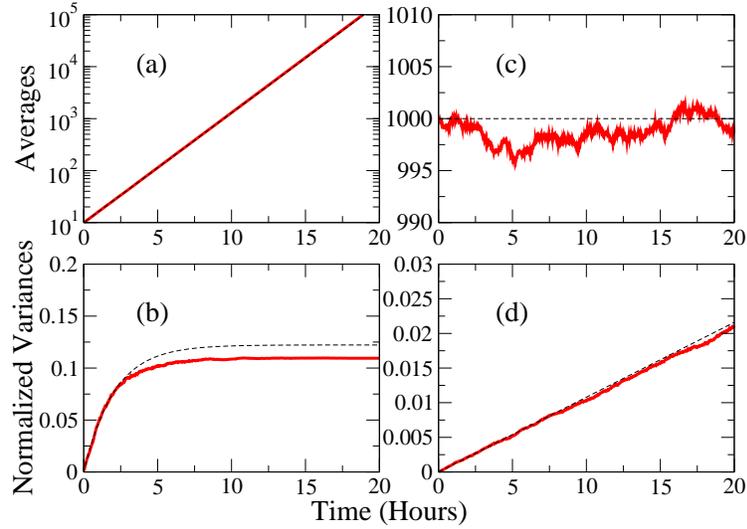}
\caption{\label{fig4} (Color online) Time-evolution of 
the normalized variance of a birth-death process of lysogens 
when the system size increases (a,b), or when it remains constant (c,d). 
(a) and (b) are time-evolutions of the mean and the normalized variance of 
lysogens when the system size increases exponentially in time, 
$r=0.54$ and $\delta=0.054$.
(c) and (d) depict those of lysogens when their growth and lysis rates 
are the same, $r=\delta=0.54$. 
Solid lines represent the results of stochastic simulations 
while dotted lines are the results of the macroscopic equation (a and c)
or the results of the linear Fokker-Plank equation (b and d).
}
\end{center}
\end{figure}

{\bf CASE 1:} $r>\delta$. When the growth rate is greater than the lysis rate, the
system size is increasing in time and the second moment of
$\xi_{I_1}$ evolves in time according to the solution of
Eq.~(\ref{BD-moments}): 
$\langle \xi^{2}_{I_1}(t) \rangle=
\frac{(r+\delta)}{(r-\delta)}i_1(0)
e^{2(r-\delta)t}(1-e^{-(r-\delta)t})$. Then the normalized
variance of lysogens reads
\begin{equation}
\label{BD-const}
\frac{\langle \delta I^{2}_{1}(t) \rangle}{\langle I_1(t)
\rangle^{2}} =\frac{\Omega_o \langle
\xi^{2}_{I_1}(t)\rangle}{\langle I_1(t)
\rangle^{2}}=\frac{(r+\delta)}{(r-\delta)I_1(0)}(1-e^{-(r-\delta)t})
\end{equation}
Asymptotically the normalized variance approaches a constant value
$(r+\delta)/((r-\delta)I_1(0))$, in good agreement with the results 
of stochastic simulations (see Fig.~\ref{fig4}(b)).

{\bf CASE 2:} $r=\delta$. When the growth rate is the same as the lysis rate, 
the system size remains constant and the
normalized variance increases linearly in time: 
$\langle \delta I^{2}_{1}(t) \rangle / \langle I_1(t) \rangle^{2}
=(r+\delta)t/I_{1}(0)$, exactly reproduced by stochastic simulations 
as shown in Fig.~\ref{fig4}(d).

\subsection{Complete Infection System: the Dynamics of Covariances of
Stochastic Fluctuations}

In this subsection, we discuss the effects of noise on phage-mediated competition. 
We explore the dynamical patterns of the normalized variances and covariances of the complete infection 
system, from which we identify the major source of stochastic fluctuations and assess their
magnitude. In the complete infection system, all bacteria in strain 1 are lysogens and 
all bacteria in strain 2 are susceptible to phage infection. 
Bacterial strain 1 (lysogens), while decoupled from the rest of the system, 
play a role as the source of the phage, triggering a massive infection process in
the susceptible bacterial strain 2. 
Throughout this subsection, we make pair-wise comparisons between the results of the deterministic
equations, stochastic simulations and of the linear Fokker-Plank equation.

\begin{figure}
\begin{center}
\includegraphics[width=10cm]{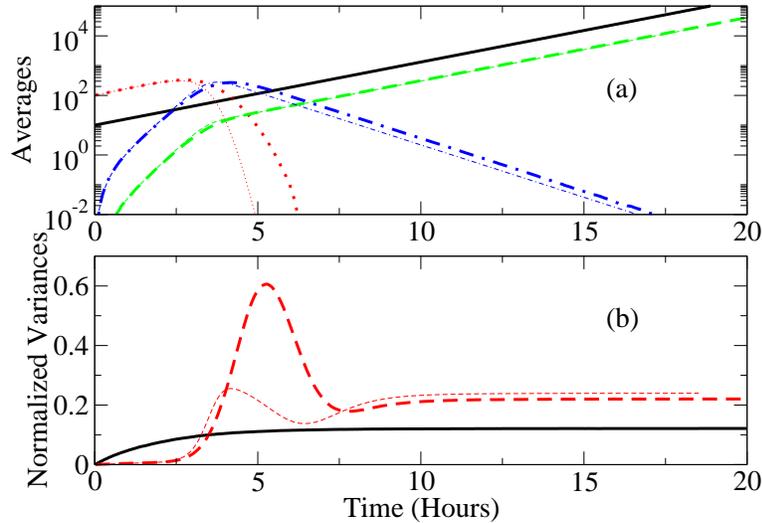}
\caption{\label{fig5} (Color online) Time-evolution of the mean
values of bacterial subpopulations (a) and the
normalized variance of total population of bacterial strain 1 and
2 (b). (a) Each subpopulation is represented by two lines; thick lines 
come from macroscopic equations in Eq~(\ref{macroscopic}) and thin lines
are obtained from stochastic simulations. The four bacterial subpopulations 
are represented by different line patterns: bacterial strain 1 in lysogenic state 
(solid lines), bacterial strain 2 in susceptible (dotted lines), 
lysogenic (dashed lines) and latent (dot-dashed lines) states.
(b) Thick solid and dashed lines represent the normalized variances 
of the bacterial strain 1 and 2 from stochastic simulations, respectively, while 
thin solid and dashed lines denote those from the linear Fokker-plank equation,
respectively. The initial condition is $I_1(0)=10$, $S_2(0)=100$ and the rest are zero. 
The parameter values are $\delta=0.054$, $\lambda=0.81$, $\kappa_2=0.00054$, $P_2=0.98$.}
\end{center}
\end{figure}

Fig.~\ref{fig5}(a) shows the time evolution of bacterial populations 
in the susceptible, lysogenic and latent states.
While bacteria of strain 1 (lysogens) grow exponentially unaffected by phage, 
the susceptible bacteria of strain 2 undergo a rapid infection process, 
being converted either into a latent state or into lysogens. 
The number of bacteria in the latent state increases, reaches a peak at a later stage 
of infection process, and then decays exponentially at a rate $\lambda$. 
As time elapses, eventually all susceptible bacteria are depleted 
from the system and both bacterial strains become lysogens, 
which grow at a net growth rate $a-\delta$. 
The ratio of the two bacterial strains (lysogens) remains unchanged
asymptotically.
Note that although the initial population size of bacterial strain 1 
is one-tenth of the initial population size of bacterial strain 2, 
strain 1 will outnumber strain 2 at a later time due to phage-mediated competition.
Pair-wise comparisons between the results from stochastic simulation and
those from deterministic equations are made in Fig.\ref{fig5}(a). 
They agree nicely with each other except a noticeable discrepancy found in 
the population size of susceptible bacteria.

The temporal patterns of the normalized variances of the two bacterial strains 
are illustrated in Fig.~\ref{fig5}(b). 
The normalized variance of bacterial strain 1 (lysogens) 
increases logistically while that of bacterial strain 2 increases logistically 
for the first few hours and then rapidly rises to its peak upon the onset of a massive 
phage infection process taking place in the susceptible bacterial strain 2.  
Asymptotically, susceptible bacteria are depleted from the system and all remaining
bacteria 
are lysogens, and their normalized variances converge to a constant as given by Eq.~(\ref{BD-const}).
The results from stochastic simulations indicate that the magnitude of noise, defined 
as the ratio of the standard deviation to the average value, of bacterial strain 2 
reaches a maximum value, 80$\%$, during the time interval while the number of susceptible bacteria 
dramatically drops and the number of latent bacteria begins to decay from its peak value.
This suggests that the stochastic fluctuations in the phage-mediated competition mainly
come from the stochastic dynamics of the susceptible bacteria undergoing infection 
process and death.
Note that the linear Fokker-Plank equation underestimates the peak value of the normalized
variance, compared to the stochastic simulations, 
while the stationary values of the normalized variances of both bacterial strains from two methods 
agree nicely.

\begin{figure}
\begin{center}
\includegraphics[width=15cm]{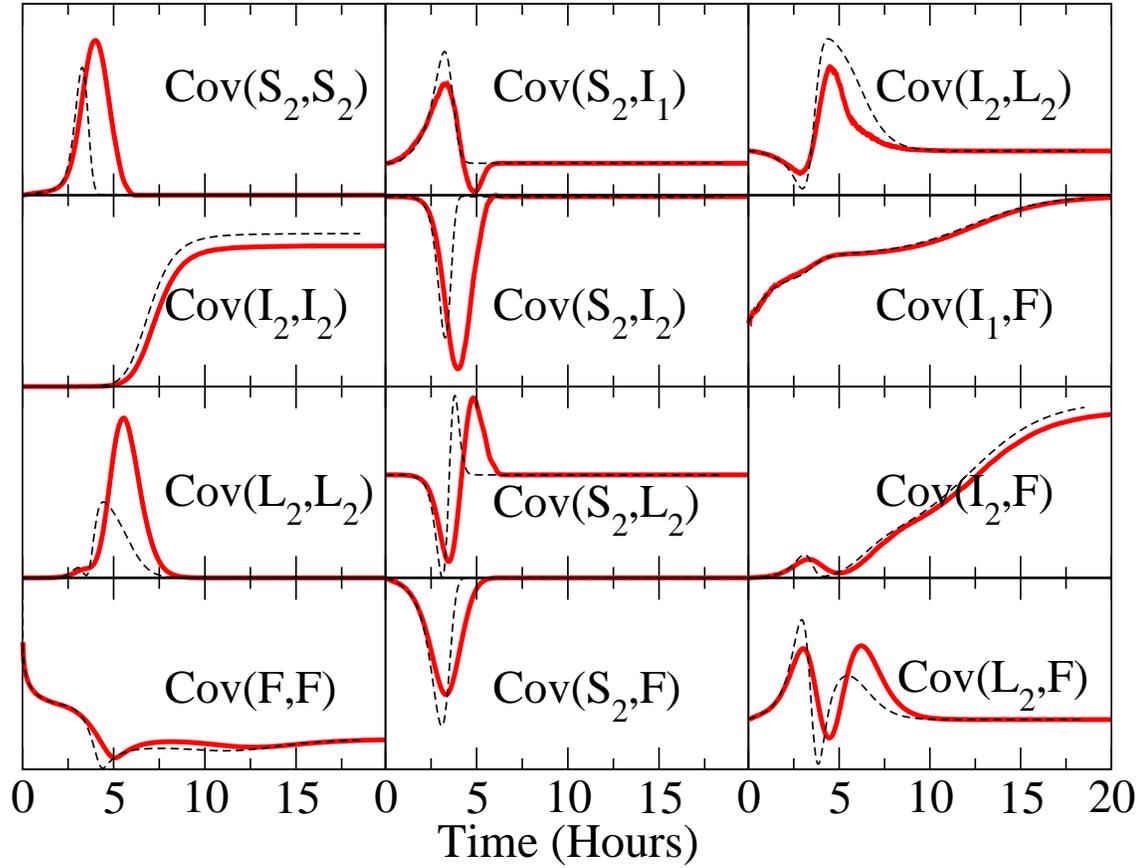}
\caption{\label{fig6} (Color online) Time evolution of the normalized (co-)variances 
of bacterial populations in different states. 
See the main text for a formal definition of 
the normalized covariance $Cov(\alpha_i,\beta_j)$. 
The time-evolution of each co-variance is plotted with two lines: solid lines are
from stochastic simulations while dashed lines are from the linear 
Fokker-Plank equation Eq.~(\ref{second_moment_FP}). 
``F'' stands for $\Phi$. 
The same parameters and initial conditions are used as in
Fig.~\ref{fig5}. Only 12 out of 15 co-variances are plotted.}
\end{center}
\end{figure}

Fig.~\ref{fig6} shows the dynamical patterns of the normalized covariances of 
bacterial populations. 
We utilize the normalized covariances to identify the main source of stochastic fluctuations 
in phage-mediated competition.
The normalized covariance of the total population of bacterial strain 2, $Var(N_2)$, 
is composed of the 6 normalized covariances of the subpopulations of bacterial strain
2, $Cov(S_2,S_2)$, $Cov(I_2,I_2)$, $Cov(L_2,L_2)$, $Cov(S_2,I_2)$, $Cov(S_2,L_2)$ and
$Cov(I_2,L_2)$.
The peak values of $Cov(S_2,I_2)$, $Cov(S_2,L_2)$ and $Cov(I_2,L_2)$ are much smaller (ten
times smaller for this particular choice of parameters) than those of 
$Cov(S_2,S_2)$, $Cov(I_2,I_2)$ and $Cov(L_2,L_2)$.
The normalized covariance of $Cov(L_2,L_2)$ reaches its peak value, 
the largest value among all normalized covariances, at the exact moment when the normalized
variance of the total population of bacterial strain 2, $Var(N_2)$, hits its maximum value
as shown in Fig.~\ref{fig5}(b). 

\noindent
This indicates that the stochastic fluctuations in the phage-mediated 
competition mainly come from the fluctuations of the bacterial population in the latent state. 
Those fluctuations originate from two events: 
incoming population flow from the just infected susceptible bacteria into the latent bacterial 
population and outgoing population flow by infection-induced lysis of 
the bacteria in the latent state. 
This indicates that the magnitude of noise does depend on the values of 
the kinetic parameters (an infection causing cintact rate $\kappa_j$ and infection-induced 
lysis rate $\lambda$) of the complete infection system, and
this also suggests the possibility of large deviations from the deterministic invasion 
criterion due to stochastic noise. 
The time evolution of the normalized co-variances that are obtained 
from both a linear Fokker-Plank equation and 
stochastic simulations agrees to each other nicely. 
This agreement validates the applicability of van Kampen's $\Omega_o$-expansion method 
to a nonlinear stochastic system which grows indefinitely.

\subsection{The Effect of Stochastic Noise on the Invasion Criterion}

\begin{figure}
\begin{center}
\includegraphics[width=10cm]{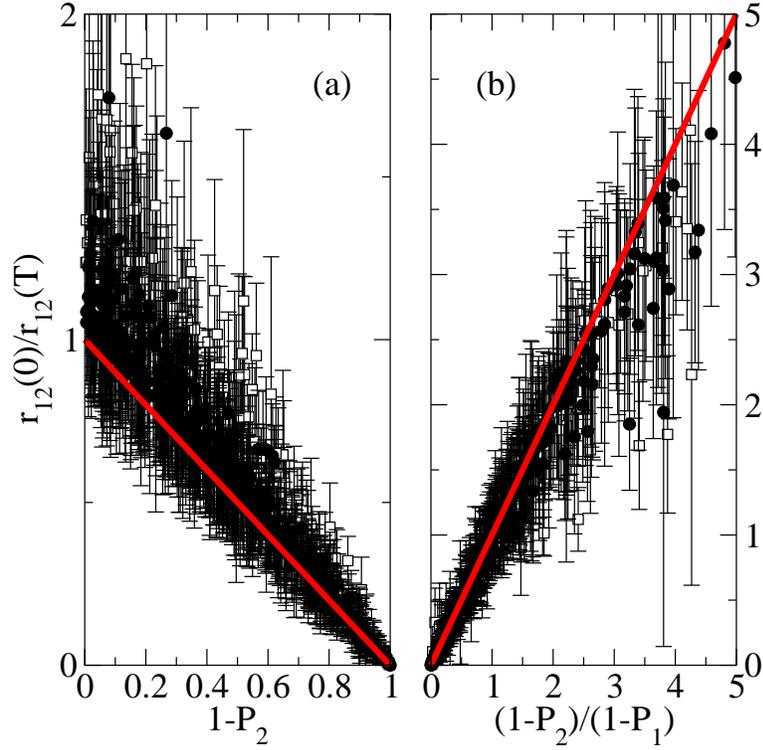}
\caption{\label{fig7} (Color online) Verification of the invasion criterion by means of 
stochastic simulations: (a) a complete infection system case where 
bacterial strain 1 is lysogen and only bacterial strain 2 is susceptible, 
(b) a general infection system where both strains are susceptible to phage. 
Thick red lines represent the invasion criterion obtained from deterministic
equations, i.e., $r_{12}(0)/r_{12}(T)=(1-P_2)/(1-P_1)$ where the time $T$
is chosen to be a sufficiently long time so that there are no more susceptible bacteria in the system. 
Error bars are the standard deviations calculated from the stochastic simulation. 
Filled circles are for a fast infection process ($\kappa > 10\kappa_o$,$\lambda > 10 \lambda_o$)
while open squares are for slow infection ($\kappa_o <\kappa <10
\kappa_o$,$\lambda_o < \lambda < 10 \lambda_o$). 
Each one of about 500 data points in each figure represents the result of stochastic simulations, 
averaged over $10^{4}$ realizations. Please see the main text for the choice of parameter values.
}
\end{center}
\end{figure}

In this subsection we investigate the effects of noise on the validity of the invasion criterion 
and measure the deviations of the stochastic results from the simple relationship 
in Eq.~(\ref{invasion}) obtained from the deterministic model.
For further analysis of the effect of noise on phage-mediated competition,
we need to perform stochastic simulations with different values of kinetic parameters and 
to investigate the effect of noise on the invasion criterion. 
We consider both a complete infection system having
only lysogens in bacterial strain 1 ($P_1=0$) in Fig.\ref{fig7}(a) and a
generalized infection system in Fig.\ref{fig7}(b) where both strains are susceptible to phage infection, 
yet with different degrees of susceptibility and vulnerability to phage. 
The invasion criterion obtained from the deterministic equations is expressed
with a simple relationship between the initial and final ratios of population 
sizes of two strains and phage pathologies: 
$r_{12}(0)/r_{12}(T)=(1-P_2)/(1-P_1)$.
Here $T$ is defined as a sufficiently long time such that there are no more 
susceptible bacteria left to undergo the infection process and only
lysogens are in the system. 
To amplify the effect of noise on phage-mediated competition, 
we set the initial sizes of bacterial populations to be small; 
they are randomly chosen from an interval $10< S_{j}(0), I_{j}(0)<110$.
To make sure that the complete infection system reaches a stationary state of having only lysogens within 24 hours, 
we limit the values of the infection-causing contact rate $\kappa_j$ 
and of the infection-induced lysis rate $\lambda$: 
$\kappa_j>\kappa_o$ and $\lambda>\lambda_o$ where $\kappa_o=0.000054$ and $\lambda_o=0.081$.
We distinguish infection processes based on their speed: a very fast
infection process ($\kappa > 10\kappa_o$, $\lambda > 10 \lambda_o$) and a slow infection process 
($\kappa_o<\kappa<10 \kappa_o$, $\lambda_o < \lambda < 10 \lambda_o$). 
The values of all other kinetic parameters in Fig.~\ref{fig2} are
randomly selected from the biologically relevant intervals (see Table 1):
$0< \delta < 0.108$, $1< \chi <100$ and $0< P_j<1$.
For about 500 sets of parameters for each figure in Fig.~\ref{fig7}, 
we measure the average and the standard deviation of the ratio
$r_{12}(0)/r_{12}(T)$ after taking ensemble average over $10^{4}$ realizations.
Note that the standard deviation is measured as a deviation from the macroscopic (true)
value and it is not normalized by the square root of the sampling size. 
We find that the average values of the ratios $r_{12}(0)/r_{12}(T)$ 
still fall onto the linear relationship with phage pathologies, 
independently of other kinetic parameters. 
However, the ratios of $r_{12}(0)/r_{12}(T)$ are broadly distributed 
around the mean value with large deviations, especially when the phage is more
pathological on strain 2, i.e., as $P_2 \rightarrow 1$ with a fixed $P_1=0$ 
for Fig.~\ref{fig7}(a) and when the phage is more pathological on strain 1 than 
on bacterial strain 2, i.e., $(1-P_2)/(1-P_1)>>1$ or $P_1>>P_2$ for Fig.~\ref{fig7}(b)
\footnote[4]{The apparent contradiction between these two cases is a result of differences
in the initial 
condition as well as the fundamental differences in partial and complete resistance.}. 
Thus the probabilistic model of phage-mediated competition in bacteria confirms that the quantitative amount of
phage-mediated competition can be still predictable despite inherent stochastic
fluctuations, yet deviations can be also large, depending on the values of phage
pathologies.

\section{CONCLUSION}

We utilized a probabilistic model of a phage-mediated invasion process to investigate the
conjecture that (i) a bacterial community structure is shaped by phage-mediated
competition between bacteria, and to examine (ii) the effect of intrinsic noise 
on the conclusions obtained from a deterministic model of the equivalent
system. The system under our consideration consists of two strains of bacteria: both
bacterial strains are susceptible to phage infection and one invasive bacterial strain
contains lysogens carrying the prophage. 
Two bacterial strains are genetically identical except in their susceptibilities to phage and
in phage pathologies on them. We restricted the infection system such that
bacteria grow in a log phase, i.e., there is no resource competition between them.

Despite the historical success of deterministic models of ecological processes,   they
produce, at best, only partially correct pictures of stochastic processes in ecological
systems. A good number of examples of the  failures of deterministic models in
ecology are presented in Ref.~\cite{durrett:1994}. The principal flaw of deterministic
models is their reliance on many, sometimes unphysical, assumptions such as 
continuous variables, complete mixing and no rare events. 
Thus, we used both Fokker-Plank equations and stochastic
simulations in the study of stochastic phage-mediated invasion processes in bacteria.
Van Kampen's system size expansion$^{\cite{vankampen:2001}}$ was used to obtain the
linear
Fokker-Plank equation while the Gillespie algorithm was used for stochastic simulations.
We found that the linear Fokker-Plank equation is a good approximation to the nonlinear
dynamics of the stochastic phage-mediated invasion process; the time evolutions of
co-variances of bacterial populations from both Fokker-Plank equation and stochastic
simulations agree well with each other.

To investigate the role of noise during phage-mediated processes, we measured the magnitude
of noise, defined as the ratio of the standard deviation of bacterial population to its mean
as time elapses. After a sufficiently long time, compared to the typical time scale of
infection processes, all surviving bacteria are lysogens, which undergo the process of
growth and spontaneous lysis. As it is a simple birth-death process with a positive net
growth rate, the magnitude of noise asymptotically converges to a rather small constant
value. However, it was found from both the linear Fokker-Plank equation and stochastic simulations
that the magnitude of noise of the bacterial subpopulations both in the susceptible and 
latent states rapidly increases and reaches a peak value in the middle of the massive phage-induced lysis event. Thus the
population size of the susceptible and latent bacteria is subject to large deviations from its mean.

We investigated the effect of noise on the invasion criterion, which is defined as the
condition of the system parameters for which the invading bacterial strain harboring and
transmitting the phage takes over the ecological niches occupied by bacterial strains
susceptible to the phage. In our previous work$^{\cite{joo:2005}}$, we showed, by using 
{\it in vitro} experiments and deterministic models, that phage-conferred competitive
advantage could be quantitatively measured and is predicted and that the final ratio
$r_{12}(T)$ of population sizes of two competing bacteria is determined by only two 
quantities, the initial ratio $r_{12}(0)$ and the phage pathology (phage-induced
mortality), independently of
other kinetic parameters such as the infection-causing contact rate, the spontaneous and
infection-induced lysis rates, and the phage burst size. 
Here we found from stochastic simulations that the average values
of the ratios $r_{12}(0)/r_{12}(T)$ still fall onto the deterministic linear
relationship with phage pathologies, independently of other kinetic parameters.
However, the ratios $r_{12}(0)/r_{12}(T)$ are broadly
distributed around the mean value, with prominently large deviations when the phage is
more pathological on the invading bacterial strain than the strain 2, i.e., $P_1>>P_2$. 
Thus the probabilistic model of
phage-mediated competition in bacteria confirms that the quantitative amount of
phage-mediated competition can still be predictable despite inherent stochastic
fluctuations, yet deviations can also be large, depending on the values of phage
pathologies.

Here we assumed that the bacterial growth rates and lysis rates are identical 
in the two strains. Relaxing this assumption has a drastic simplifying effect as 
the steady state is determined solely by the net growth rates of the two strains.
Regardless of initial conditions in the generalized infection system, all bacteria that
survive after a massive phage infection
process are lysogens, so long as the phage-infection is in action on both bacterial strains. 
If the net growth rates of two
strains are such that $r_1-\delta_1>r_2-\delta_2>0$, asymptotically bacterial strain 1
will outnumber strain 2, regardless of phage pathologies and initial population sizes. If
the
net growth rate of any bacterial strain is negative, it will go extinct. Thus the
non-trivial case is only when the growth rates of two bacterial strains are identical.

We significantly simplified many aspects of complex pathogen-mediated
dynamical systems to obtain this stochastic model. The two most
prominent yet neglected features are the spatial distribution and the
connectivity pattern of the host population. As demonstrated by stochastic
contact processes on complex networks (e.g., infinite scale-free networks)
or on d-dimensional hypercubic
lattices$^{\cite{newman:2002,liggett:1999,pastorsatorras:2001,andjel:1996,durrett:1991,kuulasmaa:1982}}$,
these two effects may dramatically change the dynamics and stationary
states of the pathogen-mediated dynamical systems. While our experimental
system does not necessitate incorporation of spatial effects, complete
models of real pathogen-modulated ecological processes, e.g.,
phage-mediated competition as a driving force of the oscillation of two V.
cholera bacterial strains, one toxic (phage-sensitive) and the other
non-toxic (phage-carrying and resistant)$^{\cite{faruque:2005}}$, 
may need to take these effects into account.

\newpage

\begin{appendix}

\noindent
{\bf APPENDIX A: SYSTEMATIC EXPANSION OF THE MASTER EQUATION}
\\
In this appendix we provide the result of the systematic expansion of the master equation in Eq.~(\ref{master}).
The master equation in the new variables reads
\begin{eqnarray}
\label{expansion}
&&\frac{\partial \Pi}{\partial t}
-\sum_{\alpha=S_1,S_2,I_1,I_2,L_1,L_2,\Phi} \Bigl \{
\Omega^{\frac{1}{2}}_o \frac{\partial \alpha}{dt} \frac{\partial \Pi}{\partial \xi_{\alpha}}
\Bigr \}
\\ \nonumber
&&= \sum_j \biggl \{
a \sum_{\alpha=S_j,I_j} \Bigl \{
(-\Omega^{-\frac{1}{2}}_o \frac{\partial}{\partial \xi_\alpha}
+\frac{\Omega^{-1}_o}{2} \frac{\partial^{2}} {\partial \xi_\alpha^{2}} -...)
(\Omega_o \alpha +\Omega^{\frac{1}{2}}_o \xi_{\alpha}) \Pi
\Bigr \}
\\ \nonumber
&&+\kappa_j \Bigl \{
(1+\Omega^{-\frac{1}{2}}_o \frac{\partial}{\partial \xi_\Phi}
+\frac{\Omega^{-1}_o}{2} \frac{\partial^{2}} {\partial \xi_\Phi^{2}}+...)
(1+\Omega^{-\frac{1}{2}}_o \frac{\partial}{\partial \xi_{S_{j}}}
+\frac{\Omega^{-1}_o}{2} \frac{\partial^{2}} {\partial \xi_{S_{j}}^{2}}+...)
\\ \nonumber
&&
\Bigl [
(1-P_j)(1-\Omega^{-\frac{1}{2}}_o \frac{\partial}{\partial \xi_{I_j}}
+\frac{\Omega^{-1}_o}{2} \frac{\partial^{2}} {\partial \xi_{I_j}^{2}} -...)+
P_j (1-\Omega^{-\frac{1}{2}}_o \frac{\partial}{\partial \xi_{L_j}}
+\frac{\Omega^{-1}_o}{2} \frac{\partial^{2}} {\partial \xi_{L_j}^{2}} -...)
\Bigr ]
-1
\Bigr \}
\\ \nonumber
&& (\Omega_o \phi +\Omega^{\frac{1}{2}}_o \xi_{\Phi})
(\Omega s_j +\Omega^{\frac{1}{2}}_o \xi_{S_j}) \Pi
+\delta \Bigl \{
(1+\Omega^{-\frac{1}{2}}_o \frac{\partial}{\partial \xi_{I_j}}
+\frac{\Omega^{-1}_o}{2} \frac{\partial^{2}} {\partial \xi_{I_j}^{2}}+...)
\\ \nonumber
&& (1-\Omega^{-\frac{1}{2}}_o \frac{\partial}{\partial \xi_\Phi}
+\frac{\Omega^{-1}}{2} \frac{\partial^{2}} {\partial \xi_\Phi^{2}}+...)^{\chi}
-1 \Bigr \}
(\Omega_o i_j +\Omega^{\frac{1}{2}}_o \xi_{I_j})  \Pi
\\ \nonumber
&&+\lambda \Bigl \{
(1+\Omega^{-\frac{1}{2}}_o \frac{\partial}{\partial \xi_{L_j}}
+\frac{\Omega^{-1}_o}{2} \frac{\partial^{2}} {\partial \xi_{L_j}^{2}}+...)
(1-\Omega^{-\frac{1}{2}}_o \frac{\partial}{\partial \xi_\Phi}
+\frac{\Omega^{-1}_o}{2} \frac{\partial^{2}} {\partial \xi_\Phi^{2}}+...)^{\chi}
-1 \Bigr \}
\\ \nonumber
&& (\Omega_o l_j +\Omega^{\frac{1}{2}}_o \xi_{L_j})  \Pi
\biggr \}
\nonumber
\end{eqnarray}

\bigskip

\noindent
{\bf APPENDOX B: LINEAR FOKKER-PLANK EQUATION DERIVED FROM SYSTEMATIC EXPANSION OF THE MASTER EQUATION}
\\
From Eq.~(\ref{expansion}) we can collect terms of order $\Omega^{0}$ and obtain
the linear Fokker Plank equation, 
\begin{eqnarray}
\label{FP}
&&\frac{\partial \Pi}{\partial t}=\sum_j \biggl \{
a(-\frac{\partial \xi_{S_j}\Pi}{\partial \xi_{S_j}}
-\frac{\partial \xi_{I_j}\Pi}{\partial \xi_{I_j}}
+\frac{s_j}{2}\frac{\partial \Pi}{\partial \xi_{S_j}^{2}}
+\frac{i_j}{2}\frac{\partial \Pi}{\partial \xi_{I_j}^{2}})
\\ \nonumber
&&+\kappa_j \Omega_o \phi s_j \Bigl \{
\frac{\partial^{2}}{\partial \xi_\Phi \partial \xi_{S_j}}
+\frac{1}{2}(\frac{\partial^{2}}{\partial \xi_{\Phi}^{2}}
+\frac{\partial^{2}}{\partial \xi_{S_j}^{2}})
+(1-P_j)(\frac{1}{2}\frac{\partial^{2}}{\partial \xi_{I_j}^{2}}
-\frac{\partial^{2}}{\partial \xi_\Phi \partial \xi_{I_j}}
-\frac{\partial^{2}}{\partial \xi_{S_j} \partial \xi_{I_j}}
)
\\ \nonumber
&&+P_j(\frac{1}{2}\frac{\partial^{2}}{\partial \xi_{L_j}^{2}}
-\frac{\partial^{2}} {\partial \xi_\Phi \partial \xi_{L_j}}
-\frac{\partial^{2}} {\partial \xi_{S_j} \partial \xi_{L_j}}
)
\Bigr \} \Pi
+\kappa_j \Omega_o \Bigl \{
\frac{\partial}{\partial \xi_\Phi}+
\frac{\partial}{\partial \xi_{S_j}}
-(1-P_j)\frac{\partial}{\partial \xi_{I_j}}
-P_j\frac{\partial}{\partial \xi_{L_j}}
\Bigr \}
\\ \nonumber
&&(\phi \xi_{S_j} \Pi+ s_j \xi_\Phi \Pi)
+\delta \Bigl \{
\frac{\partial \xi_{I_j} \Pi}{\partial \xi_{I_j}}
-\chi \frac{\partial \xi_{I_j} \Pi}{\partial \xi_{\Phi}}
+\frac{i_j}{2}\frac{\partial^{2} \Pi}{\partial \xi_{I_j}^{2}}
+i_j(\frac{\chi}{2}+{}_{\chi}C_{2}) \frac{\partial^{2} \Pi}{\partial \xi_{\Phi}^{2}}
-\chi i_j \frac{\partial^{2} \Pi}{\partial \xi_{I_j} \partial \xi_{\Phi}}
\Bigr \}
\\ \nonumber
&&+\lambda \Bigl \{
\frac{\partial \xi_{L_j} \Pi}{\partial \xi_{L_j}}
-\chi \frac{\partial \xi_{L_j} \Pi}{\partial \xi_{\Phi}}
+\frac{l_j}{2}\frac{\partial^{2} \Pi}{\partial \xi_{L_j}^{2}}
+l_j(\frac{\chi}{2}+{}_{\chi}C_{2})\frac{\partial^{2} \Pi}{\partial \xi_{\Phi}^{2}}
-\chi l_j \frac{\partial^{2} \Pi}{\partial \xi_{L_j} \partial \xi_{\Phi}}
\Bigr \}
\biggr \}
\end{eqnarray}

We obtain the first moments of the Gaussian noise by multiplying the Eq.~(\ref{FP}) by $\xi_{\alpha}$ and 
integrating over all $\underline{\xi}$, i.e., $\int \xi_{\alpha} d \Pi=
\langle \xi_{\alpha} \rangle$.
\begin{eqnarray}
\label{first_moment_FP}
\frac{d \langle \xi_{S_j} \rangle  }{dt}&=& a \langle \xi_{S_j} \rangle
-\kappa_j \Omega_o (\phi \langle \xi_{S_j} \rangle + s_j \langle \xi_{\Phi} \rangle)
\\ \nonumber
\frac{d \langle \xi_{I_j} \rangle  }{dt}&=& (a-\delta) \langle \xi_{I_j} \rangle
+\kappa_j \Omega_o (1-P_j)(\phi \langle \xi_{S_j} \rangle +s_j \langle \xi_{\Phi} \rangle)
\\ \nonumber
\frac{d \langle \xi_{L_j} \rangle  }{dt}&=&
\kappa_j \Omega_o P_j(\phi \langle \xi_{S_j} \rangle +s_j \langle \xi_{\Phi} \rangle)
-\lambda \langle \xi_{L_j} \rangle
\\ \nonumber
\frac{d \langle \xi_{\Phi} \rangle  }{dt}&=&
\sum_j \Bigl \{
\delta \chi \langle \xi_{I_j} \rangle +\lambda \chi \langle \xi_{L_j} \rangle
-\kappa_j \Omega_o \phi (\langle \xi_{S_j} \rangle + s_j \langle \xi_{\Phi} \rangle)
\Bigr \}
\nonumber
\end{eqnarray}

Similarly we obtain the second moments (covariance) of the Gaussian noise 
by multiplying the Eq.~(\ref{FP}) by $\xi_{\alpha} \xi_{\beta}$ and 
integrating over all $\underline{\xi}$, i.e., 
$\int \xi_{\alpha} \xi_{\beta} d \Pi=\langle \xi_{\alpha} \xi_{\beta} \rangle$.

\begin{eqnarray}
\label{second_moment_FP}
\frac{d \langle \xi_{S_j} \xi_{S_{j'}} \rangle}{dt}&=&
2a \langle \xi_{S_j} \xi_{S_{j'}} \rangle
+(a s_j+ \kappa_j \Omega_o s_j \phi) \delta_{j j'}
-\Bigl \{
\kappa_j \Omega_o (\langle \xi_{S_j} \xi_{S_{j'}} \rangle \phi
\\ \nonumber
&&+\langle \xi_{S_{j'}} \xi_{\Phi} \rangle s_j)
+(j \longleftrightarrow j') \Bigr \}
\\ \nonumber
\frac{d \langle \xi_{I_j} \xi_{I_{j'}} \rangle}{dt}&=&
2a \langle \xi_{I_j} \xi_{I_{j'}} \rangle
+(a i_j+ \delta i_j+k_j \Omega_o \phi s_j (1-P_j)) \delta_{j j'}
-2 \delta \langle \xi_{I_j} \xi_{I_{j'}} \rangle
\\ \nonumber
&&+\Bigl \{
\kappa_j \Omega_o (1-P_j)(\langle \xi_{S_j} \xi_{I_{j'}} \rangle \phi
+\langle \xi_{I_{j'}} \xi_{\Phi} \rangle s_j)
+(j \longleftrightarrow j') \Bigr \}
\\ \nonumber
\frac{d \langle \xi_{L_j} \xi_{L_{j'}} \rangle}{dt}&=&
(\lambda l_j +k_j \Omega_o \phi s_j P_j)\delta_{j j'}
-2 \lambda \langle \xi_{L_j} \xi_{L_{j'}} \rangle
+\Bigl \{
\kappa_j \Omega_o P_j (\langle \xi_{S_j} \xi_{L_{j'}} \rangle \phi
+\langle \xi_{L_{j'}} \xi_{\Phi} \rangle s_j)
\\ \nonumber
&&+(j \longleftrightarrow j') \Bigr \}
\\ \nonumber
\frac{d \langle \xi_{\Phi}^{2} \rangle}{dt}&=&
\sum_j \Bigl \{
\kappa_j \Omega_o s_j \phi-2 \kappa_j \Omega_o (\langle \xi_{S_j} \xi_{\Phi} \rangle \phi
+\langle \xi_{\Phi}^{2} \rangle s_j)
+2 \chi (\delta \langle \xi_{I_j} \xi_{\Phi} \rangle
+\lambda \langle \xi_{L_j} \xi_{\Phi} \rangle )
\\ \nonumber
&&+(\chi+2 {}_{\chi}C_{2})(\delta i_j+\lambda l_j)
\Bigl \}
\\ \nonumber
\frac{d \langle \xi_{S_j} \xi_{I_{j'}} \rangle}{dt}&=&
2a \langle \xi_{S_j} \xi_{I_{j'}} \rangle
-(1-P_j) \kappa_j \Omega_o s_j \phi \delta_{j,j'}
-\kappa_j \Omega_o (\langle \xi_{S_j} \xi_{I_{j'}} \rangle \phi
+\langle \xi_{I_{j'}} \xi_{\Phi} \rangle s_j)
\\ \nonumber
&&+(1-P_{j'}) \kappa_{j'} \Omega_o (\langle \xi_{S_j} \xi_{S_{j'}} \rangle \phi
+\langle \xi_{S_j} \xi_{\Phi} \rangle s_{j'})
-\delta \langle \xi_{S_j} \xi_{I_{j'}} \rangle
\\ \nonumber
\frac{d \langle \xi_{S_j} \xi_{L_{j'}} \rangle}{dt}&=&
a \langle \xi_{S_j} \xi_{L_{j'}} \rangle
-P_j \kappa_j \Omega_o s_j \phi \delta_{j,j'}
-\lambda \langle \xi_{S_j} \xi_{L_{j'}} \rangle
-\kappa_j \Omega_o (\langle \xi_{S_j} \xi_{L_j'} \rangle \phi
+\langle \xi_{L_{j'}} \xi_{\Phi} \rangle s_j)
\\ \nonumber
&&+P_{j'} \kappa_{j'} \Omega_o (\langle \xi_{S_j} \xi_{S_{j'}} \rangle \phi
+\langle \xi_{S_j} \xi_{\Phi} \rangle s_{j'})
\\ \nonumber
\end{eqnarray}

\begin{eqnarray}
\frac{d \langle \xi_{S_j} \xi_{\Phi} \rangle}{dt}&=&
a \langle \xi_{S_j} \xi_{\Phi} \rangle
+\kappa_j \Omega_o s_j \phi
+\chi \sum_{j'} (\delta \langle \xi_{S_j} \xi_{I_{j'}} \rangle
+\lambda \langle \xi_{S_j} \xi_{L_{j'}} \rangle)
\\ \nonumber
&&-\sum_{j'} (\kappa_{j'} \Omega_o \langle \xi_{S_{j'}} \xi_{S_j} \rangle
\phi
+\kappa_{j'} \Omega_o \langle \xi_{S_j} \xi_{\Phi} \rangle s_{j'})
-\kappa_j \Omega_o (\langle \xi_{S_{j}} \xi_{\Phi} \rangle \phi
+\langle \xi_{\Phi}^{2} \rangle s_j)
\\ \nonumber
\frac{d \langle \xi_{I_j} \xi_{L_{j'}} \rangle}{dt}&=&
(a-\delta) \langle \xi_{I_j} \xi_{L_{j'}} \rangle
-\lambda \langle \xi_{I_j} \xi_{L_{j'}} \rangle
+\kappa_j \Omega (1-P_j)
( \langle \xi_{S_j} \xi_{L_{j'}} \rangle \phi
+\langle \xi_{L_{j'}} \xi_{\Phi} \rangle s_j )
\\ \nonumber
&&+\kappa_{j'} \Omega_o P_{j'}
( \langle \xi_{I_j} \xi_{S_{j'}} \rangle \phi
+\langle \xi_{I_j} \xi_{\Phi} \rangle s_{j'})
\\ \nonumber
\frac{d \langle \xi_{I_j} \xi_{\Phi} \rangle}{dt}&=&
(a-\delta) \langle \xi_{I_j} \xi_{\Phi} \rangle
-\delta \chi i_{j}
+\chi \sum_{j'} (\lambda \langle \xi_{I_j} \xi_{L_{j'}} \rangle
+\delta \langle \xi_{I_j} \xi_{I_{j'}} \rangle )
-\kappa_j \Omega_o s_j \phi (1-P_j)
\\ \nonumber
&&-\sum_{j'} \kappa_{j'} \Omega_o (\langle \xi_{S_{j'}} \xi_{I_{j}} \rangle \phi
+\langle \xi_{I_{j}} \xi_{\Phi} \rangle s_{j'})
+\kappa_j \Omega_o (1-P_j) (\langle \xi_{S_j} \xi_{\Phi} \rangle \phi
+\langle \xi_{\Phi}^{2} \rangle s_j)
\\ \nonumber
\frac{d \langle \xi_{L_j} \xi_{\Phi} \rangle}{dt}&=&
-P_j \kappa_j \Omega_o s_j \phi
-\lambda \chi l_j
-\lambda \langle \xi_{\Phi} \xi_{L_{j}} \rangle
+\chi \sum_{j'} (\delta \langle \xi_{L_{j}} \xi_{I_{j'}} \rangle
+\lambda \langle \xi_{L_j} \xi_{L_{j'}} \rangle )
\\ \nonumber
&&- \sum_{j'} \kappa_{j'} \Omega_o (\langle \xi_{S_{j'}} \xi_{L_{j}} \rangle \phi
+\langle \xi_{L_{j}} \xi_{\Phi} \rangle s_{j'})
+\kappa_j \Omega_o P_j (\langle \xi_{S_j} \xi_{\Phi} \rangle \phi
+\langle \xi_{\Phi}^{2} \rangle s_j)
\nonumber
\end{eqnarray}
\end{appendix}

\bigskip

\noindent
{\bf APPENDIX C: STOCHASTIC BIRTH-DEATH PROCESSESS}
\\ 
In this section we present an exact solution of the master equation
of a stochastic birth-death process, i.e., a prototype of
all birth-death systems which consists of a population of non-negative 
integer individuals $X$ that can occur with a $x$ population size
~\cite{vankampen:2001,gardiner:2004}. 
The concept of birth and death is usually that only a finite number of $X$ 
are born and die at a given time. The transition probabilities can be written 
\begin{equation}
T(x'|x;t)=t^{+}(x)\delta_{x',x+1}+t^{-}(x)\delta_{x',x-1}.
\end{equation}
Thus there are two processes: birth, $x \rightarrow x+1$, with a 
transition probability $t^{+}(x)$, and death, $x \rightarrow x-1$, 
with a transition probability $t^{-}(x)$.
The master equation then takes the form, 
\begin{equation}
\frac{d P_t(x)}{dt}=(E-1)t^{-}(x)P_t(x)+(E^{-1}-1)t^{+}(x)P_t(x)
\end{equation}
where $E$ is a step operator, e.g., $E^{\pm1}f(x)=f(x\pm1)$.
This expression remains valid at boundary points if we impose $t^{+}(0)=t^{-}(-1)=0$.

In the case of the growth and spontaneous lysis process of 
bacteria carrying prophage, $\xrightarrow{r} X \xrightarrow{\delta}$ with 
$t^{+}(x)=rx$ and $t^{-}(x)=\delta x$, the master equation
takes the simple form
\begin{equation}
\label{masterBD}
\frac{dP_t(x)}{dt}=(E^{-1}-1) r x P_t(x)+(E^{+1}-1) \delta x P_t(x)
\end{equation}
To solve Eq.~(\ref{masterBD}), we introduce the generating function 
$G(s,t)=\sum^{\infty}_{x=0} s^{x} P_t(x)$ so that 
\begin{equation}
\label{GFBD}
\partial_{t}G(s,t)=f(s) \partial_{s} G(s,t)
\end{equation}
where $f(s)=(rs-\delta)(s-1)$. 
We find a substitution that provides the desirable transformation of 
variable, $f(s) \partial_{s}=f(s)\frac{\partial z}{\partial s}
\frac{\partial}{\partial z}=-\partial_{z}$, 
\begin{equation}
z=-\int\frac{ds}{f(s)}=\frac{1}{\delta-r} 
log \Bigl ( \frac{s-1}{rs-\delta} \Bigr )
\end{equation}
This substitution, $G(s,t)=\psi(z,t)$, gives
\begin{equation}
\partial_{t}\psi(z,t)+\partial_{z}\psi(z,t)=0 
\end{equation}
whose solution is an arbitrary function of $t-z$. 
We write the solution of the above equation as $\psi(z,t)=F[e^{-t+z}]$, 
so
\begin{equation}
G(s,t)=F \Bigl [ e^{-t} \Bigl ( \frac{s-1}{r s-\delta} 
\Bigr )^{\frac{1}{\delta-r}} \Bigr ]
\end{equation}
Normalization requires $G(1,t)=\sum_x P_t(x)=1$, and hence $F(0)=1$. 
The initial condition $P_{t=0}(x)=\delta_{x,x_o}$ determines $F$, 
which means 
\begin{equation}
G(s,0)=s^{x_o}=F \Bigl [ \Bigl (\frac{s-1}{r s-\delta}
\Bigr )^{\frac{1}{\delta-r}} \Bigr ]
\end{equation}
so that 
\begin{equation}
\label{GF}
G(s,t)=\Bigl ( \frac{rs-\delta+\delta (1-s) e^{-t(\delta-r)}}
{rs-\delta+r(1-s) e^{-t(\delta-r)}} \Bigr )^{x_o}
\end{equation}
Eq.~(\ref{GF}) can be expanded in a power series in $s$ to
produce the conditional probability density $P_t(x) \equiv P(x,t|x_o,0)$,
which is the complete solution of the master equation Eq.~(\ref{masterBD}). 
Because it is of little practical use and complicated, 
we do not present the conditional probability density here, 
but compute the moment equations from the generating function in Eq.~(\ref{GF})
\begin{eqnarray}
\Bigl [ \frac{\partial log G(s,t)}{\partial s} \Bigr ]_{s=1}&=&\langle x(t) \rangle
\\ \nonumber
\Bigl [\frac{\partial^{2} log G(s,t)}{\partial s^{2}} \Bigr ]_{s=1}&=&
\langle x(t)^{2} \rangle-\langle x(t) \rangle^{2} -\langle x(t) \rangle,
\end{eqnarray}
obtaining
\begin{eqnarray}
\langle x(t) \rangle &=& x_o e^{t(r-\delta)}
\\ \nonumber
\langle \delta x(t)^{2} \rangle &=& 
\langle x(t)^{2} \rangle-\langle x(t)\rangle^{2}
=x_o \frac{r+\delta}{r-\delta} 
\bigl ( e^{2t(r-\delta)}-e^{t(r-\delta} \bigr ), 
\end{eqnarray}
which exactly corresponds to the the mean and the variance from the
linear Fokker-Plank equation.



\bigskip

\noindent
{\bf ACKNOWLEDGMENT}\\
This work was supported by a Sloan Research Fellowhship to R.A. 
and by NIH grant 5-R01-A1053075-02 to E.H.

\begin{table}[htbp]
\begin{center}
\caption{\label{table1} Parameters used for the numerical
simulation of the phage-mediated competition in {\it B.
bronchiseptica}. 
The two undetermined parameters $P$ and $\kappa$
[[hours$\cdot$CFU/ml]$^{-1}$] were estimated by comparing the
experimental results with those of the theoretical model and by
minimizing discrepancies.}
\begin{tabular}{cccc}
Parameter & Name & Range & Resources\\
\hline
a & (Free) growth rate & 0.54 & measured~\cite{joo:2005}
\\
$\delta$ & Spontaneous lysis rate & $0 \leq \delta < a$
& measured~\cite{joo:2005}
\\
$\lambda$ & $\phi$-induced lysis rate &
0.08 - 0.17 & measured~\cite{joo:2005}
\\
$\chi$ & Burst size & 10 - 50 & measured~\cite{joo:2005}
\\
P & Phage pathology & $0 \leq P \leq 1$ & estimated
\\
$\kappa$ & Contact rate & $\kappa>0$ & estimated
\\
$N_{max}$ & Holding capacity & $\sim 10^{9}$ & measured~\cite{joo:2005}
\\
\end{tabular}
\end{center}
\end{table}


\begin{thebibliography}{99}

\bibitem{anderson:1991} R. M. Anderson and R. M. May,
{\sl Infectious diseases of humans. Dynamics and control},
(Oxford University Press, Oxford, 1991).

\bibitem{dickmann:2002} U. Dickmann, J. A. J. Metz, M. W. Sabelis, and
K. Sigmund (Eds.) {\sl Adaptive Dynamics of Infectious Diseases: In
Pursuit of Virulence Management}, (Cambridge University Press,
Cambridge, United Kingdom, 2002).

\bibitem{hudson:1998} P. Hudson and J. Greenman,
Competition mediated by parasites: biological and theoretical progress, 
{\sl TREE} {\bf 13}:387-390 (1998) and references therein.

\bibitem{thomas:2005} F. Thomas, M. B. Bonsall and A. P. Dobson,
Parasitism, biodiversity and conservationin in F. Thomas, F. Renaud and J.
F. Guegan (Eds.) {\sl Parasitism and Ecosystems}, (Oxford University
Press, Oxford, 2005).

\bibitem{bonsall:1997} M. B. Bonsall and M. P. Hassell,
Apparent competition structures ecological assemblages, 
{\sl Nature}, {\bf 388}:371-373 (1997).

\bibitem{park:1948} T. Park, 
Experimental studies of interspecific competition I. 
Competition between populations of the flour beetle, 
{\it Tribolium confusum} and {Tribolium castaneum},
{\sl Ecol. Monogr.}, {\bf 18}:267-307 (1948).

\bibitem{holt:1994} R. D. Holt and J. H. Lawton,
The ecological consequences of shared natural enemies, 
{\sl Annu. Rev. Ecol. Syst.}, {\bf 25}:495-520 (1994).

\bibitem{holt:1985} R. D. Holt and J. Pickering, 
Infectious disease and species coexistence: a model of 
Lotka-Volterra form, 
{\sl Am. Nat.}, {\bf 125}:196-211 (1985).

\bibitem{begon:1992} M. Begon et al, 
Disease and community structure: the importance of host 
self-regulation in a host-host pathogen model, 
{\sl Am. Nat.}, {\bf 139}:1131-1150 (1992).

\bibitem{bowers:1997} R. G. Bowers and J. Turner, 
Community structure and the interplay between interspecific 
infection and competition, 
{\sl J. Theor. Biol.}, {\bf 187}:95-109 (1997).

\bibitem{greenman:1997} J. V. Greenman and P. J. Hudson,
Infected coexistence: instability with and without 
density-dependent regulation, 
{\sl J. Theor. Biol.}, {\bf 185}:345-356 (1997).

\bibitem{murray:1980} J. D. Murray, 
{\sl Mathematical Biology}, (Springer-Verlag, New York, 1980).

\bibitem{vankampen:2001} N. G. Van Kampen, {\sl Stochastic Processes 
in Physics and Chemistry}, (North-Holland, New York, 2001).

\bibitem{newman:2002} M. E. J. Newman, 
Spread of epidemic disease on networks, 
{\sl Phys. Rev. E}, {\bf 66}:016128 (2002).

\bibitem{bjornstad:2001} O. N. Bjornstad and B. T. Grenfell, 
Noisy clockwork: time series analysis of population fluctuations in
animals, 
{\sl Science}, {\bf 293}:638-643 (2001).

\bibitem{rohany:1999} P. Rohani, D. J. Earn and B. T. Grenfell, 
Opposite patterns of synchrony in sympatric disease metapopulations,
{\sl Science}, {\bf 286}:968-971 (1999).

\bibitem{grassy:2005} N. Grassly, C. Fraser and G. P. Garnett, 
Host immunity and synchronized epidemics of syphilis across the 
United States, 
{\sl Nature}, {\bf 433}:417-421 (2005).

\bibitem{liggett:1999} T. M. Liggett, {\sl Stochastic Interacting Systems: 
Contact, Voter and Exclusion Processes}, (Springer, New York, 1999).

\bibitem{goel:1971} N.S. Goel, S. C. Maitra and E. W. Montroll, 
On the Volterra and Other Nonlinear Models of Interacting Populations, 
{\sl Rev. Mod. Phys.}, {\bf 43}:231-276 (1971).

\bibitem{mckane:2005} A. J. Mckane and T. J. Newman, 
Predator-prey cycles from resonant amplification of demographic
stochasticity, q-bio.PE/0501023.

\bibitem{mcadams:1997} H. McAdams and A. P. Arkin,
Stochastic mechanisms in gene expression,  
{\sl Proc. Natl. Acad. Sci. USA}, {\bf 94}:814-819 (1997).

\bibitem{hasty:2000} J. Hasty, J. Pradines, M. Dolnik, and J. J. Collins, 
Noise-based switches and amplifiers for gene expression,
{\sl Proc. Natl. Acad. Sci. USA}, {\bf 97}:2075-2080 (2000).

\bibitem{ozbudak:2002} E. M. Ozbudak, M. Thattai, I. Kurtser, A. D.
Grossman and A. van Oudenaarden, 
Regulation of noise in the expression of a single gene, 
{\sl Nature Genetics}, {\bf 31}:69-73 (2002).

\bibitem{pedraza:2005} J. M. Pedraza and A. van Oudenaarden, 
Noise propagation in gene networks, 
{\sl Science}, {\bf 307}:1965 (2005).

\bibitem{barkai:2001} N. Barkai and S. Leibler, {\sl Nature}, 
Circadian clocks limited by noise, 
{\bf 403}:267-268 (2001). 

\bibitem{bialek:2005} W. Bialek and S. Setayeshgar, 
Physical limits to biochemical signalling, 
{\sl Proc. Natl. Acad. Sci. USA}, {\bf 102}:10040-10045 (2005).

\bibitem{vilar:2002} J. M. Vilar, H. Y. Kueh, N. Barkai, and S. Leibler, 
Mechanisms of noise-resistance in genetic oscillators,
{\sl Proc. Natl. Acad. Sci. USA}, {\bf 99}:5988-5992 (2002).

\bibitem{joo:2005} J. Joo, M. Gunny, M. Cases, P. Hudson, R. Albert, 
and E. Harvill, 
Bacteriophage-mediated competition in {\it Bordetella} bacteria, 
q-bio.PE/0507038.

\bibitem{ptashne:1992} M. Ptashne, {\sl A Genetic Switch},
(Cell Press and Blackwell Scientific Publications, Cambridge, 1992).

\bibitem{durrett:1994} R. Durrett and S. Levin, 
Importance of being discrete (and spatial), 
{\sl Theor. Pop. Biol.}, {\bf 46}:363-394 (1994). 

\bibitem{gillespie:1977} D. T. Gillespie, {\sl J. Phys. Chem.}, 
Exact stochastic simulation of coupled chemical reactions, 
{\bf 81}:2340-2361 (1977).

\bibitem{pastorsatorras:2001} R. Pastor-Satorras and A. Vespignani, 
Epidemic spreading in scale-free networks, 
{\sl Phys. Rev. Lett.}, {\bf 86}:3200-3203 (2001).

\bibitem{andjel:1996} E. Andjel and R. Schinazi,
A complete convergence theorem for an epidemic model, 
{\sl J. Appl. Probab.}, {\bf 33}:741-748 (1996).

\bibitem{durrett:1991} R. Durrett and C. Neuhauser,
Epidemics with recovery in D 2, 
{\sl Ann. Appl. Probab.}, {\bf 1}:189-206 (1991).

\bibitem{kuulasmaa:1982} K. Kuulasmaa, 
The spatial general epidemic and locally dependent random graphs, 
{\sl J. Appl. Probab.}, {\bf 19}:745-758 (1982). 

\bibitem{faruque:2005} S. M. Faruque, I. B. Naser, M. J. I. B. Islam, A.
S. G. Faruque, A. N. Ghosh, G. B. Nair, D. A. Sack, and J. J. Mekalanos, 
Seasonal epidemics of cholera inversely correlate with the prevalence of
environmental cholera phages, 
{\sl Proc. Natl. Acad. Sci.}, {\bf 102}:1702 (2005).

\bibitem{gardiner:2004} C. W. Gardiner, 
{\sl Handbook of Stochastic Methods for Physics, Chemistry and Natural Sciences},
(Springer-Verlag, Berlin, 2004).

\end{thebibliography}
\end{document}